\documentclass[pre,twocolumn,showpacs,superscriptaddress] {revtex4}
\usepackage{graphics}
\newcommand{\sinp}
{\affiliation{Theoretical Condensed Matter Physics Division, and
Centre for Applied Mathematics and Computational Science,\\
 Saha Institute of Nuclear Physics,
 1/AF Bidhannagar, Kolkata 700064, India}}
\newcommand{\cu}
{\affiliation{Department of Physics, University of Calcutta, 
92 Acharya Prafulla Chandra Road, Kolkata 700009, India}}

\begin{document}

\title
{Phase transitions in Ising model on a Euclidean network}

\author{Arnab Chatterjee} 
\email{arnab.chatterjee@saha.ac.in}
\sinp

\author{Parongama Sen}
\email{psphy@caluniv.ac.in}
\cu

%\today

\begin{abstract}
A one dimensional network on which there are
long range bonds at lattice distances $l>1$ with the probability
$P(l) \propto l^{-\delta}$ has been taken under consideration.
We investigate the critical behavior of the Ising model
on such a network where spins interact with these extra neighbours apart from
their nearest neighbours for $0 \leq \delta < 2$.
It is observed that there is a finite temperature 
phase transition in the entire range.  For $0 \leq \delta < 1$, finite size 
scaling behaviour of various quantities are consistent with mean field 
exponents while for $1\leq \delta\leq 2$, the exponents  depend on 
$\delta$.
The results are discussed in the context of 
earlier  observations  on  the topology of the 
underlying network.
\end{abstract}

\maketitle

\section{Introduction}
\label{cs:intro}
During the last few years there has been a lot of activity in the
study of networks once it was realised that networks of diverse nature 
exhibit some common features in their underlying structure. 
The Watts-Strogatz (WS) model~\cite{WS}
was proposed to simulate the small-world feature of real networks (i.e.,
 the  property that the networks have  a small diameter as well as 
 a large clustering coefficient).
In this model, the nodes are placed on a
ring and each node has connection to a finite number of nearest 
neighbours initially.
The nearest neighbour links are then  rewired with a rewiring probability $p_r$ to
form random long range links.
This  model 
displays phase transitions from a regular to a  small world to a  random network
 by varying the single parameter $p_r$ representing 
disorder. The regular to small world network
transition was shown to take place at $p_r \to 0$. 
In a slight variation of the WS model, the addition type 
network was considered in which random long range bonds are added
with a probability $p$,
keeping the nearest neighbour links undisturbed. 
Phase transition to a small world network was again  observed for
$p \to 0$~\cite{newman_crit,BA_review}.  

Phase transitions in networks can also be driven by 
factors other than the above kind of disorder.  
In many real world networks, the linking probability is 
dictated by factors like Euclidean distances separating them,  aging etc.
It is possible to achieve phase transitions 
in the network by  tuning the parameters governing such factors.
These networks are indeed not
just theoretical concepts.  
Many real world networks like the Internet at the router level, transport networks, power grid network and even collaboration networks are 
 indeed described on a Euclidean space  
in which
the geographical locations of the nodes play an important role in 
the construction of the network \cite{yook,newman,katz,collab}.
Similarly, aging factor is  important in  social networks and citation 
network \cite{BA_review,hajra,redner} where it is found that
linking with older nodes are generally less probable.  

\begin{figure}[t]
\centering{
\resizebox*{8.5cm}{!}{\includegraphics{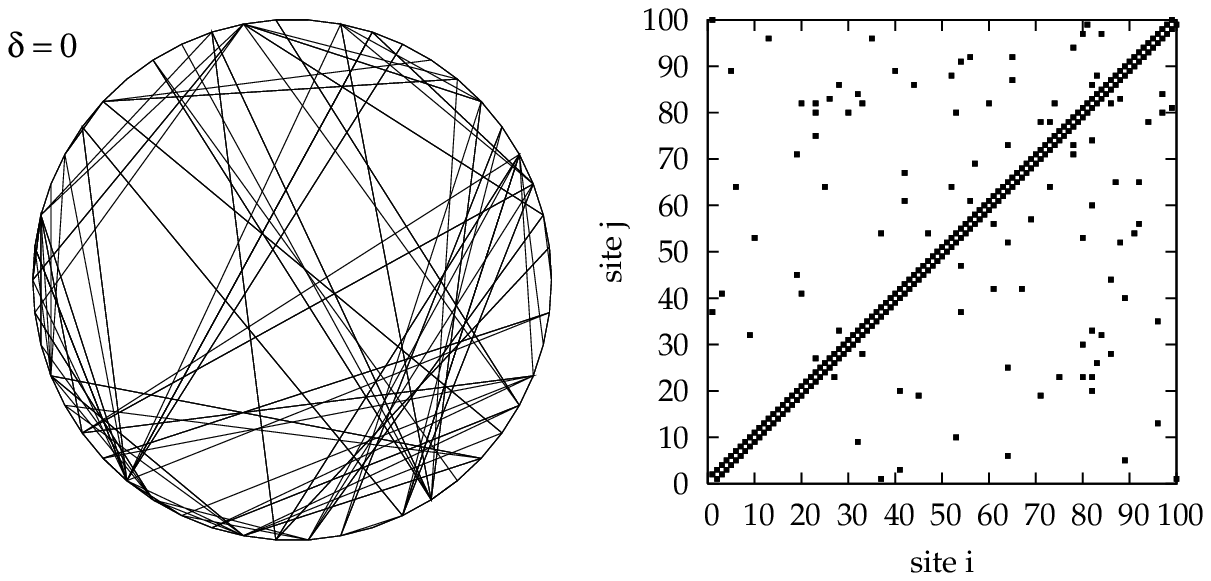}}
\resizebox*{8.5cm}{!}{\includegraphics{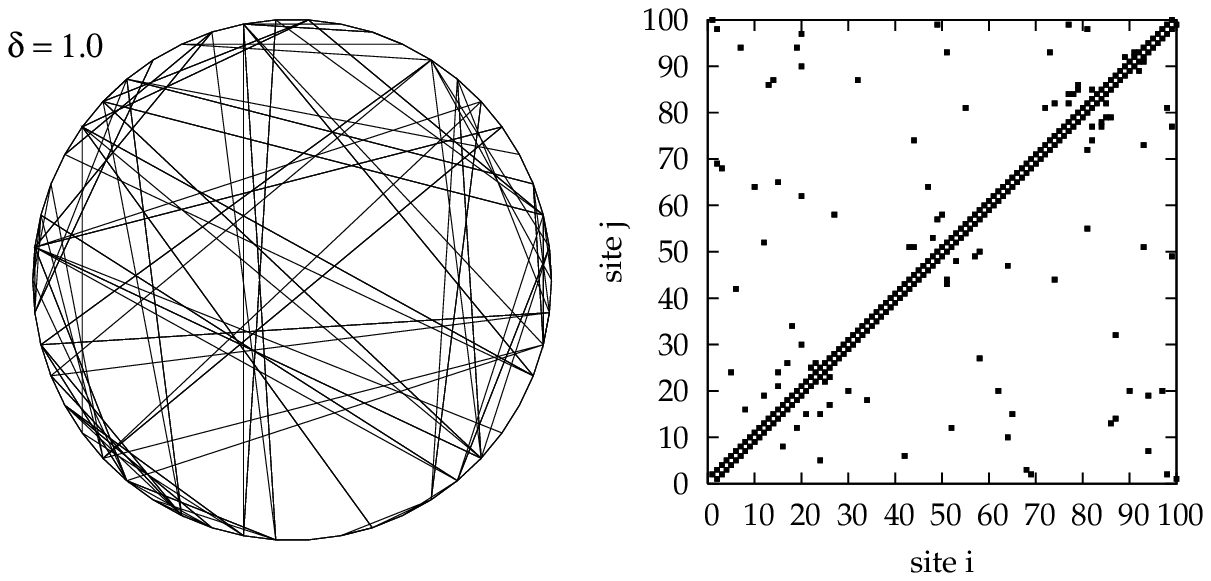}}
\resizebox*{8.5cm}{!}{\includegraphics{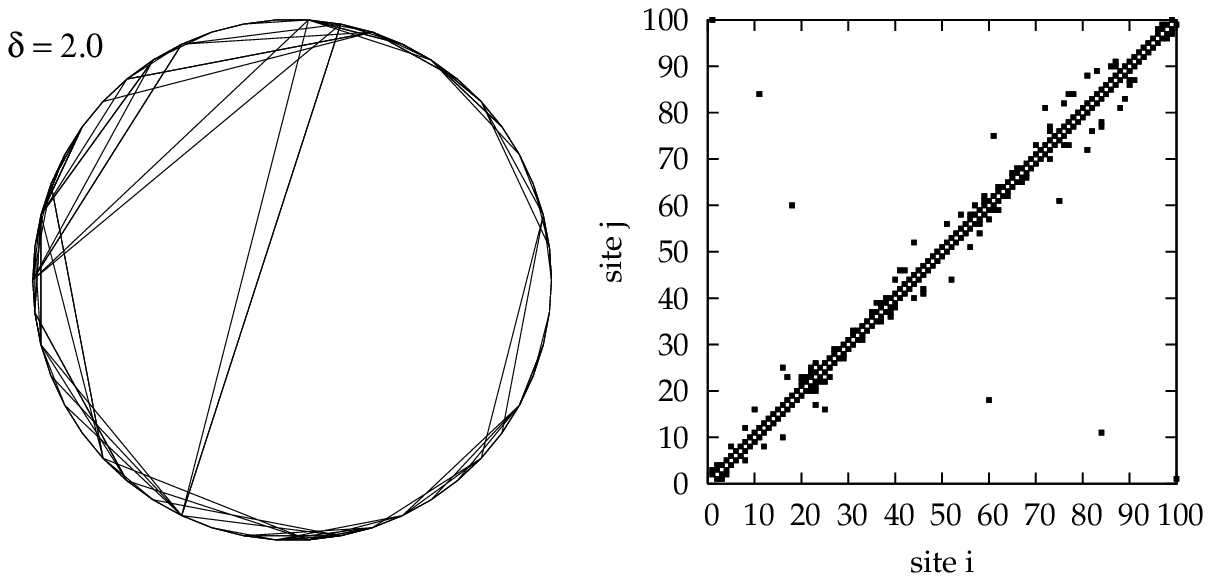}}
}
\caption{
Structures of the networks (left panel) and the corresponding adjacency
diagrams (right panel) for $\delta=0,~1.0,~2.0$ (top to bottom) in the
model for a system of size $N=100$.
}
\label{schematic}
\end{figure}

In the theoretical modelling of Euclidean networks \cite{psreview}, it is usually assumed 
that 
two nodes separated  by a Euclidean distance $l$ are connected 
with the probability $P(l)$ 
which follows a power law variation,  i.e., 
\begin{equation}
\label{eq1}
P(l) \sim l^{-\delta}.
\end{equation}
In one dimension, where the nodes are placed on a ring, the typical networks
generated for different values of $\delta$ and the corresponding
adjacency matrix  are shown graphically in Fig. \ref{schematic}.
Growing networks on Euclidean spaces where the
linking probability is modified with such a probability factor 
have also been considered recently \cite{manna_sen}.
In a real situation, however, $P(l)$ may not have such a well-defined
behaviour \cite{newman,collab}. 
 
A variation of $P(l)$ in the above form was first studied  by Kleinberg \cite{klein} 
on a two dimensional plane with the aim to find
out how navigation in the network depends on the paramater $\delta$.
Later, a number of other properties of networks (both static and dynamic) 
with such a probabilistic
attachment have been studied \cite{psreview,blumen,psbkc,mouk,biswas,zhu}.

In the present study, we have reconsidered a static one dimensional 
Euclidean network (OEN) with connection probability at 
distance $l$ given by eq~(\ref{eq1}).
We are interested in the  phase transitions which can be achieved by varying the parameter $\delta$.
Although this issue has been addressed earlier~\cite{blumen,psbkc,mouk,biswas}, there remain some questions 
about the exact nature of phase transitions in the region $0 < \delta < 2$. 
We have tried to resolve the
problems by taking a different approach here. In the earlier
studies, the topological features like the diameter, shortest paths and the
clustering coeffcient of the  network 
 had been studied.
 Here we have considered     Ising spins
 on the nodes of such a network. The critical behaviour of this system
 is expected to reflect the nature of
the network at different values of $\delta$ indirectly. 
That the Ising model undergoes phase transitions
on the addition type WS network is an established fact~\cite{ising_WS,gitterman,twenty}. 
It has also been
shown that in the small world phase the addition type WS has a mean field
nature~\cite{kim,herrero}.

In section~\ref{cs:neteusp}, we have reviewed the available results of the
Euclidean networks  and also justified the necessity of the present 
investigations. In section~\ref{cs:isnet}, studies of  Ising models 
on networks have been briefly reviewed. Our results of the study of the Ising model
on a Euclidean network is given in section~\ref{cs:iseunet}. 
Finally, in section~\ref{cs:sumcon}, summary and conclusions are given. 

\section{Network on Euclidean space} 
\label{cs:neteusp}
The one dimensional Euclidean network (OEN), with the 
connection probability  given by eq (\ref{eq1}),
is equivalent to the addition type WS model at $\delta=0$ 
when addition of long range links  can take place at any length scale. 
%Corresponding to
%the parameter $p$ in the WS model, here we have the number of random long range links 
%in the network. 
The choice of the number of  long
range neighbours in this OEN should be such that it allows a phase transition at
$\delta=0$.
In  the addition type WS model, this transition  for a system of
$N$ nodes is achieved for $p = 1/N$ (implying $p \to 0$ in
the thermodynamics limit). Connection probability $p=1/N$ implies a total number of 
$N$ long range edges in the system or the existence of 
one long range bond per node
on an average.  Therefore in order to obtain  a phase transition in the OEN, 
it is sufficient to keep $N$ long range bonds.
For large values of $\delta$ the nodes in the network 
have short range connections only and therefore it behaves as a regular network~\cite{blumen,psbkc,mouk}.
 Thus there should exist at least one phase transition in this 
network.

In all, there could be four kinds of behaviour of   this network:
(a) Regular:  when  the network is like a short ranged one dimensional system (b)
finite dimensional: when the network behaves as a system with 
effective dimensionality greater than one but still short ranged (c) Small world  and 
(d) Random. 
A few studies in which these phase transitions have been investigated
have not been able to give a concrete or unique answer. 
The network behaves
as a regular network above $\delta = 2$; this is obtained in all the earlier studies.
 The interesting region is 
$0 \leq \delta \leq 2$ as here the network  no longer behaves 
like a one-dimensional system.  In~\cite{blumen}, the behaviour of the OEN 
was studied by releasing a random walker on the network.  The results  indicated
that  the walker
has the same behaviour as on a small world network 
for all  $\delta < 2$.

To detect whether a network has small world behaviour, one can calculate
the shortest distances separating two nodes, take the average and
analyse its behaviour  with the system size. The largest of
these shortest paths is called the diameter of the network.
The average shortest path and diameter of a network are expected 
to have the same scaling behaviour.
Sen and Chakrabarti~\cite{psbkc} studied the diameters of the
OEN while the  
shortest paths were calculated by Moukarzel and de Menezes~\cite{mouk}  giving contradictory results.
In~\cite{psbkc}, it was found that the diameter behaved 
as $\ln {N}$ on chains of size $N$ 
for all $\delta < 2$ and hence it was concluded that 
	 small world behaviour occurs for $0 \leq \delta \leq 2$ while
	in~\cite{mouk}, it was argued that small world behaviour occurs only for
$\delta < 1$. In the region $1< \delta<2$, according to \cite{mouk},
the shortest distances scale as $N^{\theta}$ with  
the value of the shortest path exponent $\theta ~~ (0<\theta<1/2)$ depending on $\delta$.
The system sizes considered in~\cite{mouk} were much larger
compared to those in~\cite{psbkc} and the discrepancy of the results could be 
accounted for by this fact.
It must also be mentioned  here  the above results are  for $pN=1$, results for other values of $p$ have also
been considered in these studies.

Random and small world phases  can be 
distinguished by the clustering coefficient which remains finite
in the latter.
The estimate of the clustering coefficients of the present network 
~\cite{biswas} showed  
that below $\delta =1$ it 
vanishes. This was considered to be a signature of the network being 
random for $\delta < 1$ and based on the findings of \cite{psbkc},
it was concluded that the region $0 \leq \delta < 2$ was 
equally divided into a random (for $\delta < 1$) and a small world 
(for $\delta > 1$) phase.  
But this is by no means foolproof as in this particular 
network the clustering coefficient
 is bound to be small close to $\delta =0$  
as the number of
nearest neighbour is small when $\delta$ is small. (In the original WS model,
the clustering coefficient remained high
for finite values of $p_r \neq 1$ only when the number of nearest neighbours was at least
four to begin with.)
Thus it may not indicate a random to small world phase transition at $\delta=1$ as
conjectured in~\cite{biswas}, it
could, in principle,  correspond to the small world to a finite dimensional
network as found in~\cite{mouk}.
That there could be two transitions, 
one at $\delta = 1$ 
and the second at $\delta = 2$ was also
supported by  
simple scaling arguments for the average bond lengths \cite{biswas}.

It has been mentioned in the beginning of this section that  the
the total number of random long range bonds per site has been kept equal to $1$
on an average such that  $p=1/N$. This is also consistent with the
fact that such a choice will keep the probability normalised.
However, in an analytical approach there will be a problem at $\delta = 1$
where such  a normalisation is not possible. This issue has been discussed 
in~\cite{blumen} and does not cause a problem in numerical calculations.  

Navigation  or searching on small world 
networks are known to give rise to shortest paths which do {\it{not}}  behave as
$\ln(N)$ but rather have sublinear variations with $N$ except for special
points: this was first detected in~\cite{klein} for two dimensional lattices and
later confirmed for the WS model~\cite{moura}. In the present one dimensional 
Euclidean model this was again demonstrated in~\cite{zhu}. 
\begin{figure}[t]
\centering
\resizebox*{8.5cm}{!}{\includegraphics{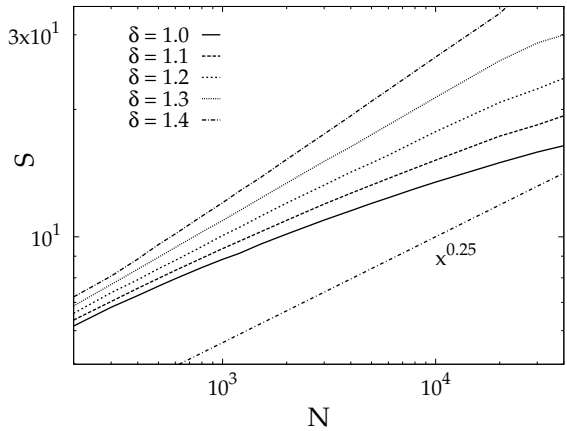}}
\caption{
Behavior of shortest path $S$ with system size $N$.
}
\label{shortpath}
\end{figure}

The more recent results~\cite{mouk,biswas} indicate that 
there is indeed a phase transition at $\delta =1$ and that  the
network behaves as a small world only for $\delta < 1$. In the present
work we want to confirm this using a completely different approach.
 If the network's
behaviour is  that of a small world 
it should be reflected in
the critical exponents of the Ising model which would assume mean field 
values. 
On the other hand, if it behaves like a finite dimensional system with effective
dimension greater than one,
a phase transition will be observed with critical exponents assuming values different
from the mean field ones (note that the lower critical dimension
of the Ising model is 1).

 Another interesting point which can be studied by considering the Ising
model on this network is the relation between the shortest path exponent 
$\theta$ and the effective dimension of the network for $1 < \delta< 2$. 
 If one interprets $\theta$ as the inverse of 
some effective dimension,
$d_{\rm eff}$, then, according to~\cite{mouk}, for some  values of $\delta > 1$,  $d_{\rm eff}$ 
is actually greater than 4. In fact we have used the burning algorithm
\cite{burn} to find out independently the shortest paths between {\it all} pairs of nodes 
in network of size $N$ and found out that indeed $\theta$  
has values  below 0.25 upto   $\delta \simeq     1.2$ (Fig. \ref{shortpath}). 
Our estimates of $\theta $ are slighlty larger than those obtained 
in~\cite{mouk}, which may be due to the smaller system size considered, but still
we find $\theta < 0.25$ over a considerable   range of values of $\delta > 1$.
The question that
naturally arises is whether the Ising model shows mean field 
behaviour here as the effective dimension is higher than the upper 
critical dimension, i.e.,  does it have a mean field behaviour even though the
network is not a small world?

\section{Ising models on networks} 
\label{cs:isnet}

Ising models with long range  neighbours have been
considered in different contexts in many earlier studies.
It is well known that the short ranged Ising model on a 
one dimensional lattice
does not show any phase transition while  interaction with all the neighbours
gives rise to a mean field like phase transition. 
The occurrence of the random long 
ranged bonds, also
enables a phase transition~\cite{ising_WS,gitterman,kim,herrero,choi,smb,lopes}.

In a analytical study of the Ising model
on a WS type network, it was assumed that
there are two nearest neighbours and additional long range neighbours
occur randomly with probability $p = \rho/N$~\cite{gitterman}
where $\rho$ is a parameter.  Phase transition occurs
here  for $\rho \geq 1$ implying that the addition of one long range bond
to each node,  (on an average)  could enhance a phase transition. 
In the numerical studies  also, the  addition
of one random long range neighbour per site  enables 
a phase transition~\cite{choi}.
The fact that the Ising model on these networks show a phase transition
even when the additional long range connections are finite in number (i.e., 
$O(N)$) 
compared to that of the 
the infinite range model (where there are $O(N^2)$ number of bonds) 
is an evidence of the key role played by 
the randomness in these networks \cite{smb}.

Another kind of long range Ising model has been considered in which
the interactions decrease with the distance as a power law, i.e,
the Hamiltonian is given by~\cite{fisher}
\begin{equation}
H = -\Sigma_{ ij } J_{ij} S_iS_j,
\end{equation}
where $J_{ij} \sim \frac {1} {r_{ij}^{\mu}}$; 
 $r_{ij}$ is the distance separating the spins at site $i$ and $j$. 
In the fully connected model, phase transitions occur at
the values $\mu = 3/2$ and $\mu=2$; for $\mu <3/2$, the
system is meanfield like while for $3/2 < \mu < 2$ it is like
a finite dimensional lattice and for $\mu > 2$, it behaves like
a one dimensional system with no phase transition.
On a  small world network, the interactions exist randomly
among neighbours, maintaining the form of $J_{ij}$ as above. However,
there is no phase transition 
displayed by this model~\cite{jeong} except at $\mu=0$ where it coincides with 
the WS model.

As mentioned earlier, the critical behaviour  of Ising model on a WS addition
type network is mean field type.
The  intriguing feature 
 is  the validity of the 
finite size scaling analysis 
in the mean field regime. 
Moreover, it has been observed~\cite{choi} that the  data collapse in the finite size scaling 
analysis can be achieved when $|T-T_c|$, the deviation from the 
critical temperature is scaled by the factor $N^{-1/\bar 
\nu}$ 
in the scaling argument with $\bar \nu = 2$.
Interpreting $\bar \nu$ as $\nu d$, where $\nu$ is the critical
exponent for the correlation function and $d$ the effective 
dimensionality of the
system, this could imply that one has effectively a system with  $d=4$ with the mean field value of $\nu = 1/2$.
This effective dimension is identical to the upper critical dimension of the Ising model.

Phase transition of the Ising model has also been
observed in     scale-free 
networks~\cite{ising_scale,doro} where the transition temperature
diverges with the system size. 
Several other aspects of Ising models on networks have been 
considered recently, for example 
the self-averaging property of Ising model on networks\cite{smb}, 
quenching dynamics \cite{pratap} etc.,
which are not directly related to the content of the present paper.

\section{Ising model on Euclidean network}
\label{cs:iseunet}

 We have considered an Ising model on a finite chain of length $N$
with periodic boundary condition.  
The nodes are assigned  position coordinates  $1,2,\ldots,N$ along the chain.
Each node is connected to its two nearest neighbours. In order to
generate the long range bonds, two  nodes are selected randomly. If $l> 1$ is the
distance separating them, they are connected with probability $P(l)$
as given by eq (1). The process is repeated  $N/2$  times generating $N$
long range bonds (each bond is counted twice so that there are $N$ long range bonds
in the system).
This ensures that there is one long range neighbour for each site on an average.
As has been mentioned in the previous section, this is sufficient to achieve a
phase transition.
For each realisation of the network, the dynamical evolution
from a uniform state (all spins up) was allowed following
a Metropolis algorithm for different temperatures $T$.

\begin{figure}%[tb]
\centering{
\resizebox*{8.5cm}{!}{\includegraphics{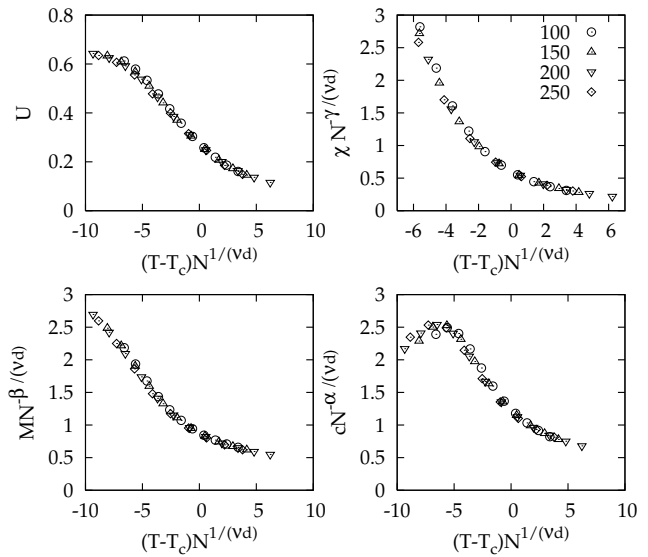}}
}
\caption{
Data collapse for different system sizes $N=100,150,200,250$
for $U$, $\chi$, $M$ and $c$, for $\delta=0.6$ ($\bar \nu = \nu d$).
}
\label{delta06all}
\end{figure}

We have computed the following quantities on this network after it reaches 
equilibrium:
\begin{itemize}
\item
{\it Magnetisation per spin~}   $m = \sum_i S_i/N$
\item
{\it Binder cumulant~}  $U = 1-\frac{\langle m^4 \rangle}{3 (\langle m^2 \rangle)^2}$.
\item
{\it Susceptibility per spin} 
is calculated from the fluctuation of the order parameter:  
\[
\chi  = \frac {N}{K_B T}[ \langle M^2 \rangle - {\langle M\rangle}^2]
\]
where $M$ is the total magnetisation,  $T$ is the temperature  
and $K_B$ the Boltzmann's constant.
\item
{\it Specific heat per spin} 
is calculated from the fluctuation of the energy of the system as
\[
c =   \frac {N}{(K_B T)^2}[ \langle E^2 \rangle - {\langle E\rangle}^2]
\]
where $E$ is the energy of the system.
\end{itemize}

\subsection{Results}
\label{cs:res}
\begin{figure}%[tb]
\centering{
\resizebox*{8.5cm}{!}{\includegraphics{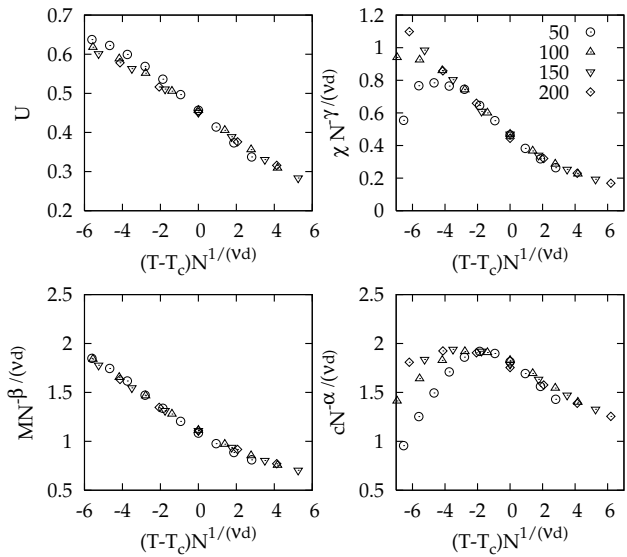}}
}
\caption{
Data collapse for different system sizes $N=50,100,150,200$
for $U$, $\chi$, $M$ and $c$, for $\delta=1.4$ ($\bar \nu = \nu d$).
}
\label{delta14all}
\end{figure}

From the intersection of $U$ for different $N$ in the plot of $U$ vs $T$ we estimate $T_c$ and a data
collapse is obtained by plotting $U$ vs $(T-T_c)N^{1/{\bar \nu}}$.
With this value of $\bar \nu$, one can now estimate the exponents $\beta $, 
$\gamma$ and $\alpha$, the critical exponents for the order parameter $m$,  
susceptibity $\chi$ and specific heat $c$ respectively  
using finite size scaling.
Figures \ref{delta06all} and \ref{delta14all} show the data collapse 
for these quantities for $\delta = 0.6$ and $\delta=1.4$. Similar results have been obtained
for other values of $\delta$. 

\subsubsection*{Transition temperature}

\begin{figure}[t]
\centering{
\resizebox*{6.5cm}{!}{\includegraphics{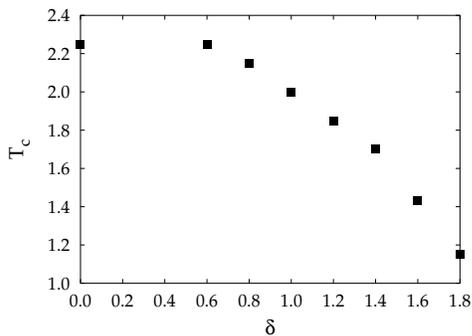}}
}
\caption{
Variation of critical temperature $T_c$ with $\delta$.
Error in this  and the next two figures are of the order
of the size of the data points.
}
\label{tcvsdelta}
\end{figure}

As expected we find a transition temperature $T_c$ 
which decreases with $\delta$ (Fig. \ref{tcvsdelta}).
$T_c$ varies very slowly in the region $0 < \delta <0.5$ and much faster for
higher values of $\delta$. 

\subsubsection*{Critical exponents}

\begin{figure}%[t]
\centering{
\resizebox*{8.5cm}{!}{\includegraphics{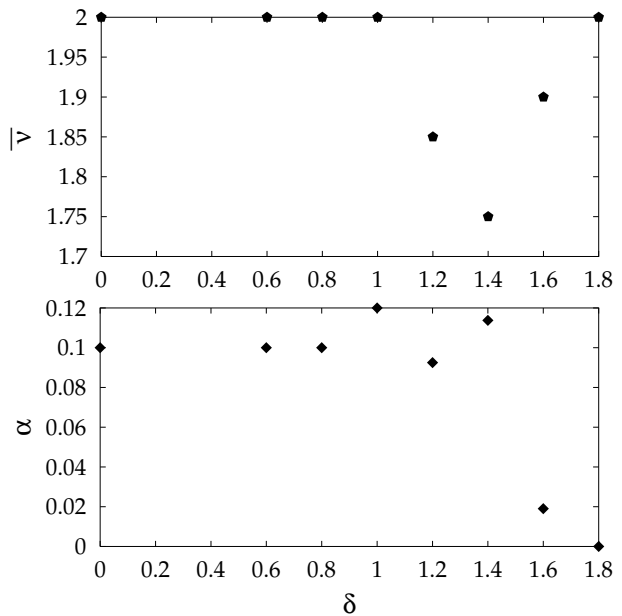}}
}
\caption{
Variation of $\bar \nu$ and specific heat exponent $\alpha$ with $\delta$.
}
\label{barnualphavsdelta}
\end{figure}

\begin{figure}%[tb]
\centering{
\resizebox*{8.5cm}{!}{\includegraphics{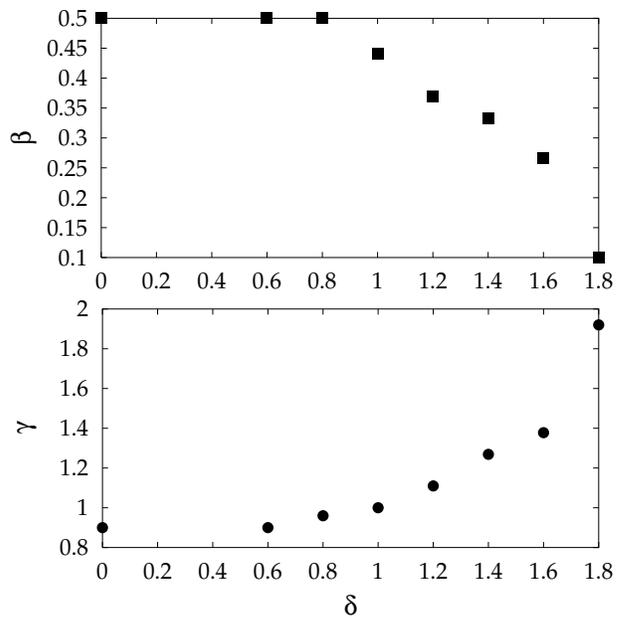}}
}
\caption{
Variation of  magnetisation exponent $\beta$ and  
susceptibility exponent $\gamma$ 
with $\delta$.
}
\label{betagammavsdelta}
\end{figure}

From the finite size scaling analysis, we obtain the critical
exponents $\bar \nu, \beta , \gamma$  and $\alpha$ (Figs \ref{barnualphavsdelta},\ref{betagammavsdelta}). 
 We find that $\bar \nu$ is equal
to 2 for the entire range $0 \leq \delta < 1$. 
However, for $\delta > 1$ we find that $\bar \nu$ decreases and then  
rises again. 


The magnetisation exponent $\beta$ is equal to 0.5 for $\delta < 1$,  
while for $\delta \geq 1$, it decreases
with $\delta$. 
The susceptibility exponent $\gamma$ is close to 1 for $\delta < 1$ and increases with $\delta$
beyond $\delta =1$. 
The specific heat exponent 
$\alpha$ remains constant at a value close to 0.1  for $\delta < 1$ and 
appears to deviate from this value as $\delta$ is made larger than $1$.  
It becomes small as $\delta$ approaches 2. 

The behaviour of all the quantities are consistent with the fact that for 
$\delta < 1$, there is a mean field behaviour of the Ising model
(consistent with the available solution for the random long-range bond
small world network \cite{lopes})
and for $\delta > 1$, its behaviour is like a finite dimensional
lattice with effective dimension greater than one. 

As has been argued for  $\delta = 0$, here also
one can interpret   $\bar \nu$ as  $\nu d$, where $d$ is the effective dimension,
 such that it satisfies $\bar \nu = 2 - \alpha$.
Since   below 
$d=4$  $\alpha$ is nonzero, $\bar \nu$  decreases from  2 at $\delta > 1$.
Theoretically, the behaviour of $\alpha$ is non-monotonic as a function of 
dimensionality $d$  as $d$  varies from 4 to 1;   
$\alpha$ is zero for both $d=4$ and $d=2$ and this
behaviour is reflected in the fact that $\bar \nu$ again rises to
2 at around $\delta = 1.8$.  However, our result for $\alpha$
does not show this non-monotonic behaviour very accurately, 
the reason possibly being that the magnitude of $\alpha$ is small ($O(0.1)$) 
even when it is non-zero and is difficult to estimate accurately
in a numerical study.

Mean field value for $\gamma$ is 1.0 which comes out to be slightly
lower in our estimate and also the value of $\alpha$ is higher than the
corresponding $0$, but the scaling relation $\alpha + 2 \beta + \gamma = 2$ is
obeyed with a high degree of accuracy. Although there are deviations of
$\alpha$ and $\gamma$  from the mean field values, the fact remains that
they show no variation over the region $0 \leq \delta < 1$, indicating that the
critical behaviour for $\delta < 1 $ is identical to that for $\delta = 0$, which is known to be mean field like.

\subsection{Effective dimension}
\label{cs:effd}
Our results confirm that the network behaves as a finite dimensional
system in the region $1 < \delta < 2$. However, 
the finite size scaling analysis 
of various quantities in the Ising model
does not  
allow the estimation of  $d$ and $\nu$  independently.
It may be expected  that the behaviour of the average shortest path
should give us some idea about the effective dimension. However, as we 
argue below, it is not possible. 
The estimate of the exact average shortest path is consistent with the
mean field picture  for $\delta < 1$ 
as it shows a logarithmic increase with $N$ (small world effect).
As mentioned before, for $\delta > 1$, it has a  power law increase $N^\theta$ 
and  we  have a region where $\theta < 0.25$. So  if
one interpretes $\theta $ as a inverse dimension, the dimensionality is 
greater than 4 here. Even then, the Ising model does not show mean field 
behaviour here. From this we conclude that although  there is an effective 
dimension coming out from the topological behaviour of the network,  
it does not exactly act as the spatial dimension as far as 
physics of Ising model is concerned. 
\begin{table}[h]
\label{tab:epsexp}
\caption{Table showing the effective dimension and comparative values of $\nu d$ from $\epsilon$ expansion (ee)  and $\bar \nu $ from finite size scaling (fss) analyses.}
\begin{tabular}{ccccc}
\hline
$\delta$ & $\nu_{\rm eff}$ & $d_{{\rm eff}}$ & $\nu d$ (ee) & $\bar \nu $ (fss)\\
\hline
1.2  & 0.56 &  3.34 &  1.85 &    1.85\\
1.4 & 0.59  &   2.90 &  1.72  &   1.75\\
1.6 & 0.61  &   2.74 &  1.66  &   1.9\\
1.8 & 0.72 &    1.36 &  0.98 &   2.0\\
\hline
\end{tabular}
\end{table}

We adopt a different approach to estimate $\nu$ and $d$ which works well
for $\delta$ close to 1.
The behaviour of the critical exponents is fully consistent with 
the fact that the
effective dimensionality $d$ decreases as $\delta $ is made larger than 1
(e.g., the value of $\gamma$ increases while that of $\beta$ decreases).
In order to find out $d$, we use the epsilon
expansions of $\alpha$, $\beta$ and $\gamma$ to first order, each of which gives
an independent estimate of $\epsilon = 4-d$: 
\begin{eqnarray}
\epsilon = 3(1-2\beta)\nonumber\\
\epsilon = 6(\gamma -1)\nonumber\\
\epsilon = 6\alpha. \nonumber
\end{eqnarray}
Ideally, all three estimates should give similar value of $\epsilon$,
but here we have used the expansions upto  linear terms only 
and the estimates may not agree very well (especially 
for the estimate from $\alpha$ and to some extent, that
from $\gamma$). It is not very convenient to
use the expansions to higher degrees also. To take care of this fact, 
we take  $\langle \epsilon \rangle$, the average  of the above three estimates 
of $\epsilon $,  and  use it 
to find out the value
of $\nu$  from  the equation
\begin{equation}
\nu = \frac {1}{2}\left(1+\frac{\langle \epsilon \rangle}{6}\right).
\end{equation}
Thus an estimate of $\nu d$ is obtained which 
can be compared to $\bar \nu$  obtained from 
the finite size scaling. We see that when $\delta -1 $ is small, 
the agreement is nice (Table 1).  
Naturally, when $\delta$ is considerably away from $1$, 
the effective dimensionality
decreases and the epsilon expansions are not much useful here,  particularly
for $\alpha$ and $\gamma$. This is reflected in the mismatch of 
the estimates  from epsilon expansion and the finite size scaling
analysis for $\delta = 1.6 $ and $\delta = 1.8$. 

\section{Summary and conclusions}
\label{cs:sumcon}
 We have investigated the behaviour of the Ising model on a one dimensional
network in which there are links to random long range neighbours existing with 
a probability which varies inversely with the distance along the original chain
following eq (\ref{eq1}).
The effective dimensionality of the network deviates from 1 continuously as $\delta$ is
made smaller than 2 and a finite temperature phase transition is observed.
The transition temperature increases as $\delta$ is made smaller and 
the critical exponents vary with $\delta$ as the dimensionality changes.
The results show that there is a mean field behaviour for $\delta < 1$ and a 
finite dimensional behaviour for $1 < \delta < 2$.
In the mean field regime, the finite size scaling analysis works with
an effective dimensionality four, the upper critical dimension of the Ising model. 
The present study confirms the conclusion made in \cite{mouk} that the small
world behaviour of the underlaying network exists for $\delta < 1$ only. 

Our results also show that the effective dimensionality cannot be simply
extracted from the behaviour of the shortest paths on this network although
the latter shows a power law behaviour with the number of nodes. 
We have used an alternative method to estimate the effective dimensionality
which works quite well for $\delta$ close to 1.

The present study was conducted to address certain issues related to the
topology of a network but the study of phase transition of Ising model
on networks has  its own inherent interest as the Ising model
has got wide applications in many interdisciplinary areas. For example, 
the study and nature of phases in such models gives us important insight
into social opinion dynamics \cite{estp} in closely knit populations, where the
networks look very similar to the models  discussed here, and 
opinions being modelled as states of spin vectors at each site.
 
\begin{acknowledgments}
Discussions with S.~M. Bhattacharjee and B.~K. Chakrabarti have been extremely helpful. PS acknowledges the financial support from CSIR grant 
no.
03(1029)/05/EMRII.
\end{acknowledgments}

\end{document}